\documentclass[aps,twocolumn,superscriptaddress,showpacs,showkeys]{revtex4}
\usepackage{graphics,graphicx,dcolumn,bm,fleqn,epic,eepic,float}
\usepackage{amssymb,amsmath,multirow,rotate,color,float,times}
\definecolor{blue}{rgb}{0,0,1}
\definecolor{red}{rgb}{1,0,0}

\begin{document}
\title{Cycles and clustering in bipartite networks}

\date{\today}

\author{Pedro G.~Lind}
\affiliation{Institute for Computational Physics, 
             Universit\"at Stuttgart, Pfaffenwaldring 27, 
             D-70569 Stuttgart, Germany}
\affiliation{Centro de F\'{\i}sica Te\'orica e Computacional, 
             Av.~Prof.~Gama Pinto 2,
             1649-003 Lisbon, Portugal}
\author{Marta C.~Gonz\'alez}
\affiliation{Institute for Computational Physics, 
             Universit\"at Stuttgart, Pfaffenwaldring 27, 
             D-70569 Stuttgart, Germany}
\author{Hans J.~Herrmann}
\affiliation{Institute for Computational Physics, 
             Universit\"at Stuttgart, Pfaffenwaldring 27, 
             D-70569 Stuttgart, Germany}
\affiliation{Departamento de F\'{\i}sica, Universidade Federal do
             Cear\'a, 60451-970 Fortaleza, Brazil}

\begin{abstract}
We investigate the clustering ability in bipartite networks where
cycles of size three are absent and therefore the standard definition
of clustering coefficient cannot be used.  
Instead, we use another coefficient given by the fraction of
cycles with size four, showing that both coefficients yield the same 
clustering properties.
The new coefficient is computed for two networks of sexual contacts, 
one monopartite and another bipartite. 
In both cases the clustering ability is similar.
Furthermore, combining both clustering coefficients we deduce an
expression for estimating cycles of larger size, which improves
previous estimations and is suitable for either monopartite and
multipartite networks. 
\end{abstract}


\maketitle


\section{Introduction} 
\label{sec:intro}

One important statistical tool to access the structure of 
complex networks arising in many systems~\cite{bollobas2,newmanrev} is
the clustering coefficient, introduced by Watts
and Strogatz~\cite{watts98} to measure ``the cliquishness of a typical
neighborhood'' in the network and given by the average fraction of
neighbors which are interconnected with each other.
This quantity has been used for instance to characterize small-world
networks~\cite{watts98}, to understand synchronization in scale-free
networks of oscillators~\cite{mcgraw05} and to characterize chemical
reactions~\cite{gleiss01} and networks of social
relationships~\cite{newman01,holme1}.  
One pair of linked neighbors corresponds to a `triangle', i.e.~a cycle
of three connections. 

While triangles may be abundant in monopartite networks, they cannot
be formed in bipartite networks~\cite{holme1,holme3,newman03}, where
two types of nodes exist and connections link only nodes of different
type. 
Thus, the standard clustering coefficient is always zero.
However, different
bipartite networks have in general different cliquishnesses and
clustering abilities~\cite{holme1}, stemming for another coefficient 
which uncovers these topological differences among bipartite networks.
Bipartite networks arise naturally in e.g.~social networks~\cite{holme3}
where the relationships (connections) depend on the gender of each
person (node), and there are situations, such as in sexual contact
networks~\cite{us1}, where one is interested in comparing clustering
properties between monopartite and bipartite compositions.

In this paper, we study the cliquishness of either monopartite and
bipartite networks, using both the standard clustering coefficient and 
an additional coefficient which gives the fraction of squares,
i.e.~cycles composed by four connections. 
As shown below, such a coefficient retains the fundamental properties
usually ascribed to the standard clustering coefficient in regular,
small-world and scale-free networks.
As a specific application, two examples of networks of sexual contacts
will be studied and compared, one being monopartite and another
bipartite.   

Furthermore, we will show that one can take triangles and squares as
the basic units of larger cycles in any network, monopartite or
multipartite.  
The frequency and distribution of larger cycles in networks have
revealed its importance in recent research for instance to
characterize local ordering in complex networks from which one is able
to give insight on their hierarchical structure~\cite{caldarelli04},
to determine equilibrium properties of specific network
models~\cite{marinari},  
to estimate the ergodicity of scale-free networks~\cite{rozenfeld},
to detect phase transitions in the topology of bosonic
networks~\cite{bianconi03}
and to help characterizing the Internet structure~\cite{bianconi04}. 
Since the computation of all cycles in arbitrarily large networks is
unfeasible, one uses approximate numerical
algorithms~\cite{rozenfeld,herrero05,yang05}  
or statistical estimations~\cite{bianconi05,barabasi05}. 
Here, we go a step further and deduce an expression to estimate 
the number of cycles of larger size, using both clustering
coefficients, which not only improves recent
estimations~\cite{barabasi05} done for monopartite networks, but at
the same time can be applied to bipartite networks and multipartite
networks of higher order. 

We start in Section \ref{sec:squares} by introducing the expression
which characterizes the cliquishness of bipartite networks, comparing
it with the usual clustering coefficient.
In Section \ref{sec:sexual} we apply both coefficients to real
networks of sexual contacts and in Section \ref{sec:new} we use them
to estimate cycles of larger size and show how it is applied to
bipartite networks.  
Conclusions are given in Section \ref{sec:discuss}.

\section{Two complementary clustering coefficients}
\label{sec:squares}

The standard definition of clustering coefficient $C_3$ is the
fraction between the number of triangles observed in one network out
from the total number of possible triangles which may appear.
For a node $i$ with a number $k_i$ of neighbors the total number of
possible triangles is just the number of pairs of neighbors
given by $k_i(k_i-1)/2$.
Thus, the clustering coefficient $C_3(i)$ for node $i$ is
\begin{equation}
C_3(i) = \frac{2t_i}{k_i(k_i-1)} . \label{c3}
\end{equation}
where $t_i$ is the number of triangles observed, i.e.~the number of
connections among the $k_i$ neighbors. 

Similarly to $C_3(i)$, a cluster coefficient $C_4(i)$ with squares
is the quotient between the number of squares and the total number
of possible squares.
For a given node $i$, the number of observed squares is given by the
number of common neighbors among its neighbors, while the total number
of possible squares is given by the sum over each pair of neighbors of
the product between their degrees, after subtracting the common node
$i$ and an additional one if they are connected.
Explicitly this clustering coefficient reads
\begin{equation}
C_4(i) = \frac{\sum_{m=1}^{k_i}\sum_{n=m+1}^{k_i}q_i(m,n)}
                {\sum_{m=1}^{k_i}\sum_{n=m+1}^{k_i} 
  \left     [a_i(m,n)+q_i(m,n)\right
  ]} , \label{c4}
\end{equation}
where $m$ and $n$ label neighbors of node $i$, $q_i(m,n)$ are the
number of common neighbors between $m$ and $n$ and 
$a_i(m,n)=(k_m-\eta_i(m,n))(k_n-\eta_i(m,n))$ with 
$\eta_i(m,n)=1+q_i(m,n)+\theta_{mn}$ and $\theta_{mn}=1$ if neighbors
$m$ and $n$ are connected with each other and $0$ otherwise. 

While $C_3(i)$ gives the probability that two neighbors of node $i$
are connected with each other, $C_4(i)$ is the probability that two
neighbors of node $i$ share a common neighbor (different from $i$).
Averaging $C_3(i)$ and $C_4(i)$ over the nodes yields two complementary 
clustering coefficients, $\langle C_3\rangle$ and $\langle
C_4\rangle$, characterizing the contribution for the network
cliquishness of the first and second neighbors respectively.
For simplicity we write henceforth $C_3$ and $C_4$ for the
  averages of $C_3(i)$ and $C_4(i)$ respectively.

\begin{figure}[htb]
\begin{center}
\includegraphics*[width=7.7cm,angle=0]{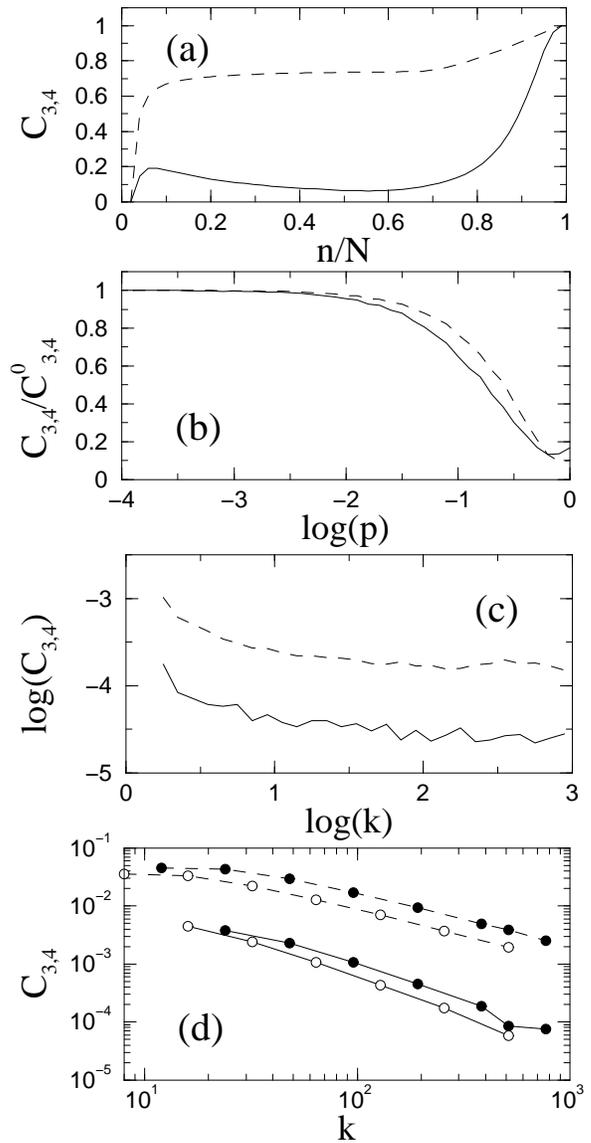}
\end{center}
\caption{\protect
         Comparisons between the standard clustering coefficient $C_3$
         in Eq.~(\ref{c3}) (dashed line) and the clustering
         coefficient $C_4$ in Eq.~(\ref{c4}) (solid line) for
         different network topologies: 
         {\bf (a)} in one regular network with $n$ neighbors symmetrically
         placed ($N=10^3$),
         {\bf (b)} in small-world networks where long-range connections
         occur with probability $p$ ($N=10^3$ and $n=4$) and
         {\bf (c)} in random scale-free networks where the distribution of the
         clustering coefficients is plotted as a function of the number
         $k$ of neighbors ($N=10^5$ and $m=2$).
         In all cases samples of $10^2$ networks were used.
         The distributions $C_3(k)$ and $C_4(k)$ are also plotted for
         {\bf (d)} Apollonian networks~\cite{hans} with $N=9844$ nodes
         ($\bullet$) and pseudo-fractal networks~\cite{dorog} with
         $N=9843$ nodes ($\circ$).}
\label{fig1}
\end{figure}

Figure \ref{fig1} shows both clustering coefficients $C_3$ and $C_4$
in several topologies. In all cases $C_3$ and $C_4$ are plotted as
dashed and solid lines respectively, and are averages over samples of
$100$ realizations.
As an example of regular networks, we use networks with
boundary conditions where each node has $n$ neighbors symmetrically
disposed. In particular, for $n=2$ one obtains a chain of nodes.
For these regular networks, Fig.~\ref{fig1}a shows the dependence of
the clustering coefficients 
on the fraction $n/N$ of neighbors, with $N=10^3$ the total number of
nodes. As one sees $C_4<C_3$ and for
either small or large fractions of neighbors both coefficients increase
abruptly with $n$. In the middle region $C_3$ is almost constant,
while $C_4$ decreases slightly.
Our simulations have shown that in regular networks
the coefficients depend only on $n/N$, i.e.~for any size of the
regular network, similar plots are obtained.

Figure \ref{fig1}b shows the coefficients for small-world networks
with $N=10^3$ nodes,
constructed from a regular network with $n=4$ neighbors symmetrically
disposed. The coefficients are computed as functions of
the probability $p$ to rewire short-range connections into long-range
connections and they are normalized as usual~\cite{watts98} to the
clustering coefficients $C_{3,4}^0$ of the underlying regular network.
As one sees, $C_4$ yields approximately the same spectrum as the
standard clustering coefficient $C_3$ being therefore able to define
the same range of $p$ for which small-world effects are
observed.
While here the small-world networks were constructed with rewiring of
short-range connections into long-range ones, the same features are
observed when using the construction procedure introduced in
Ref.~\cite{newmanproc} where instead of rewiring one uses addition of 
long-range connections.

For random scale-free networks we plot in Fig.~\ref{fig1}c the
distribution of both coefficients as functions of the number $k$ of
neighbors, using networks with $N=10^5$ nodes and by given initially
$m=2$ connections to each node.  
Here, one observes that $C_4(k)$ is almost constant as $k$
increases, reproducing the same known feature as the standard $C_3(k)$
apart a scaling factor: $C_4(k)/C_3(k)$ is approximately constant for
any $k$. 
In Fig.~\ref{fig1}d we plot the clustering distributions for two
different deterministic scale-free networks recently studied, namely
Apollonian networks~\cite{hans}, represented with bullets $\bullet$,
and pseudo-fractal networks~\cite{dorog}, represented with circles
$\circ$.
In both cases, the same power-law behavior already known for
$C_3(k)\sim k^{-\alpha}$ in these hierarchical networks is also
observed for the coefficient $C_4(k)$ with the same value of the
exponent $\alpha$.  

In short, the results shown in Fig.~\ref{fig1} give evidence that
$C_4$ is also a suitable coefficient to characterize the
topological features in several complex networks commonly done 
with the standard clustering coefficient $C_3$.
Furthermore, since $C_4$ counts squares instead of
triangles, it is particularly suited for bipartite networks.
Next, we will use this coefficient to compare different models for
networks of sexual contacts, where both monopartite and bipartite
networks arise naturally.

\section{Cycles and clustering in sexual networks}
\label{sec:sexual}

\begin{figure}[htb]
\begin{center}
\includegraphics*[width=4.28cm,angle=0]{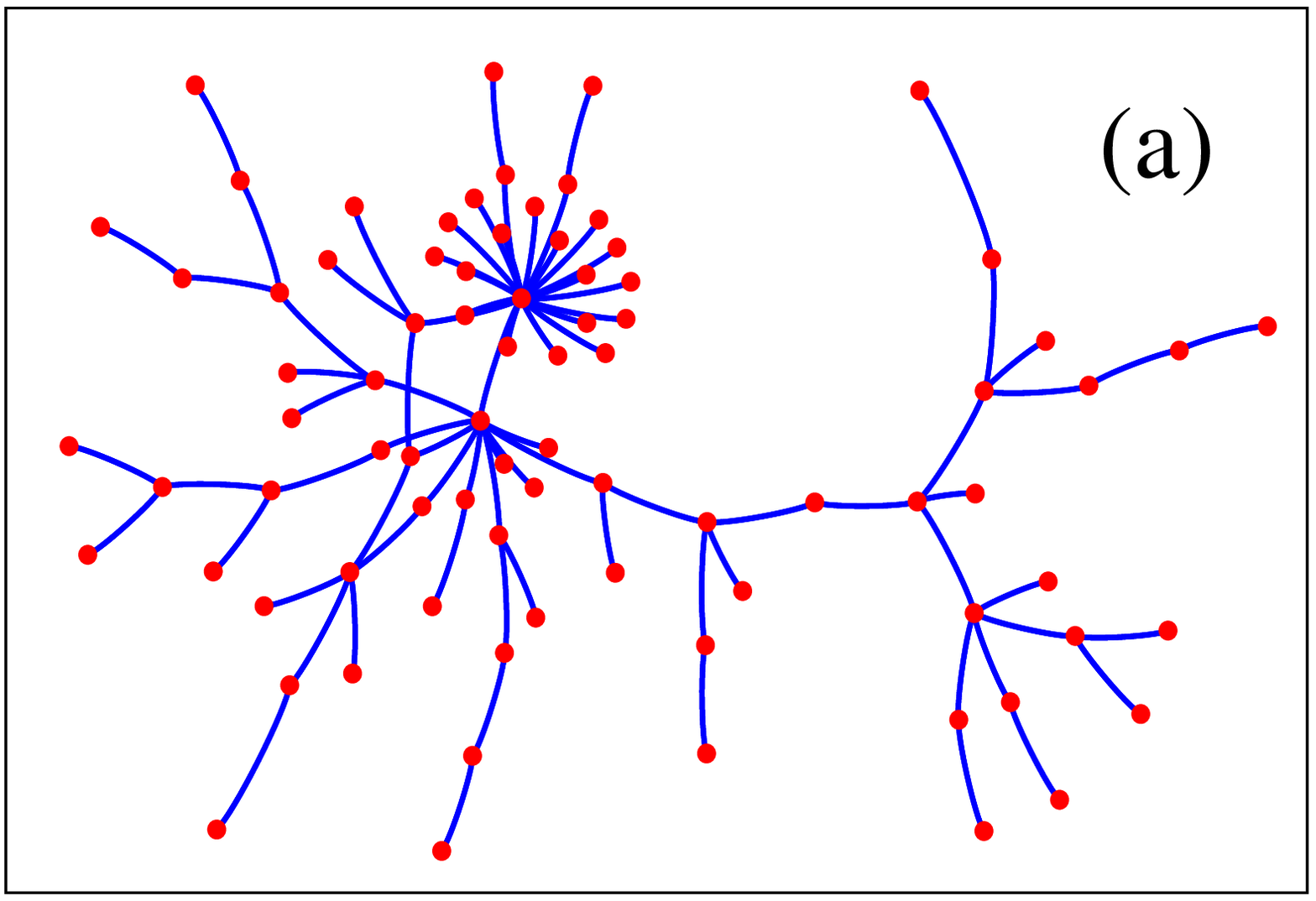}%
\includegraphics*[width=4.2cm,angle=0]{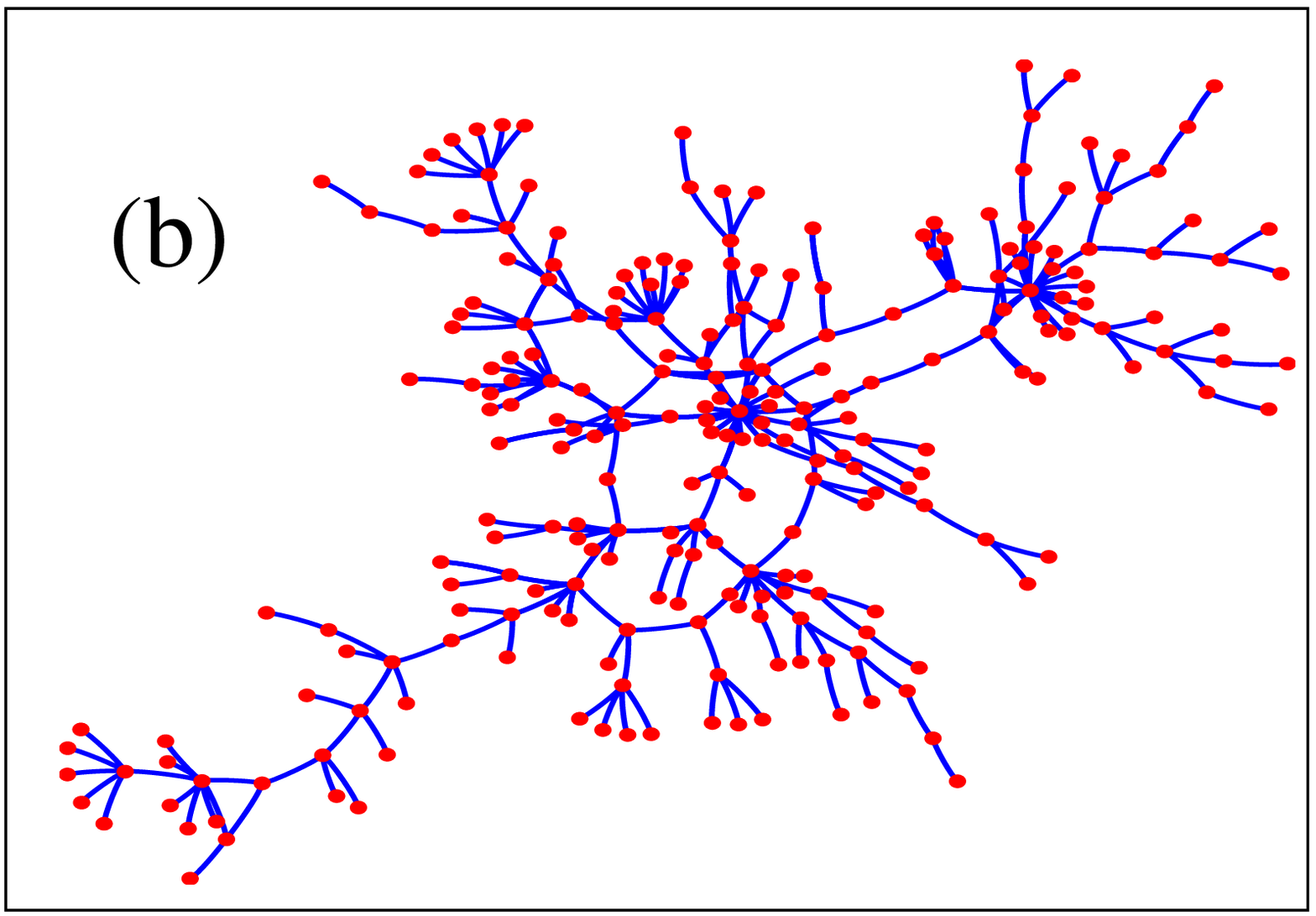}
\end{center}
\caption{\protect
         Sketch of two real sexual contact networks having
         {\bf (a)} only heterosexual contacts ($N=82$ nodes and
         $L=84$ connections) and
         {\bf (b)} homosexual contacts ($N=250$ nodes and $L=266$
         connections). 
         While in the homosexual network triangles and squares appear,
         in the heterosexual network triangles are absent (see
         Table \ref{tab1}).} 
\label{fig2}
\end{figure}

In this Section we apply both coefficients $C_3$ and $C_4$ in
Eqs.~(\ref{c3}) and (\ref{c4}) to analyze two real
networks of sexual contacts.
One network is obtained from an empirical data set, composed solely
by heterosexual contacts among $N=82$ nodes, extracted at the Cadham
Provincial Laboratory and is a 6-month block data~\cite{cospring1}
between November 1997 and May 1998. 
The other data set is the largest cluster with $N=250$ nodes in the
records of a contact tracing study~\cite{cospring2}, from 1985 to
1999, for HIV tests in Colorado Springs (USA), where most of the
registered contacts were homosexual.  
Figure \ref{fig2} sketches these two networks, where one can see that
cycles of different sizes appear. 
While the network with only heterosexual contacts is clearly
bipartite, the network with homosexual contacts is monopartite. 

For the two networks in Fig.~\ref{fig2}, Table \ref{tab1} indicates
the number $T$ of triangles, the number $Q$ of
squares and the coefficients $C_3$ and $C_4$.   
As one sees, although the 
heterosexual network has less squares than the homosexual network due to 
its smaller size, $C_4$ is much larger.
Another feature common for both neighbors is the average number of
connections per node $L/N\sim 1$.

Recently, we introduced~\cite{us1} a model to simulate the statistical
features of these networks of sexual contacts. 
The model is a sort of a granular system with low density composed by $N$ 
mobile particles representing persons and collisions between them 
representing their sexual contacts.
The collisions representing sexual contacts are governed by
dynamical rules which are carried out by means of an event-driven algorithm, 
and are based on two simple facts from sociological observations:
(i) individuals with a larger number of partners are more likely to get new 
partners and
(ii) sexual interactions do not determine the direction toward which each
agent will be moving afterward. 
Therefore, we choose a collision rule where the absolute value of the 
velocity of each agent increases with the number $k$ of sexual partners and
the moving directions after collisions are randomly selected~\cite{us1}. 
\begin{table}[htb]
\centering
\begin{tabular}{|c||c|c|c|c|c|c|}
\hline
  & $N$ & $L$ & $T$ & $Q$ & $\langle C_3\rangle$ & $\langle C_4\rangle$ \\
\hline\hline
Heterosexual & $82$ & $84$ & $0$  & $2$ & $0$ & $0.00486$ \\
\footnotesize{(Fig.~\ref{fig2}a)}  & & & & & & \\
\hline
Homosexual  & $250$ & $266$ & $11$ & $6$ & $0.02980$ & $0.00192$ \\
\footnotesize{(Fig.~\ref{fig2}b)}  & & & & & & \\
\hline\hline
Heterosexual  & $82$ & $83.63$ & $0$ & $1.45$ & $0$ & $0.01273$ \\
\footnotesize{(Agent Model)}  & & & & & & \\\hline
Homosexual  & $250$ & $287.03$ & $8.23$ & $10.52$ & $0.02302$ & $0.01224$ \\
\footnotesize{(Agent Model)}  & & & & & & \\\hline\hline
Heterosexual  & $82$ & $162$ & $0$ & $159.72$ & $0$ & $0.12859$ \\
\footnotesize{(Scale-free)}  & & & & & & \\\hline
Homosexual  & $250$ & $498$ & $45.28$ & $256.79$ & $0.08170$ & $0.02787$ \\
\footnotesize{(Scale-free)}  & & & & & & \\\hline
\end{tabular}
\caption{\protect
         Clustering coefficients and cycles in two real networks
         of sexual contacts (top), illustrated in Fig.~\ref{fig2}, one
         where all contacts are heterosexual (Fig.~\ref{fig2}a) and
         another with homosexual contacts (Fig.~\ref{fig2}b).
         In each case one indicates the values of the number $N$ of
         nodes, the number $L$ of connections, the number $T$ of
         triangles, the number $Q$ of squares and both clustering
         coefficients $C_3$ and $C_4$ in Eqs.~(\ref{c3}) and
         (\ref{c4}) respectively.
         The values of these quantities are also indicated for networks
         constructed with the agent model recently
         introduced~\cite{us1} and for random scale-free networks.}
\label{tab1}
\end{table}
\begin{figure}[htb]
\begin{center}
\includegraphics*[width=8.5cm,angle=0]{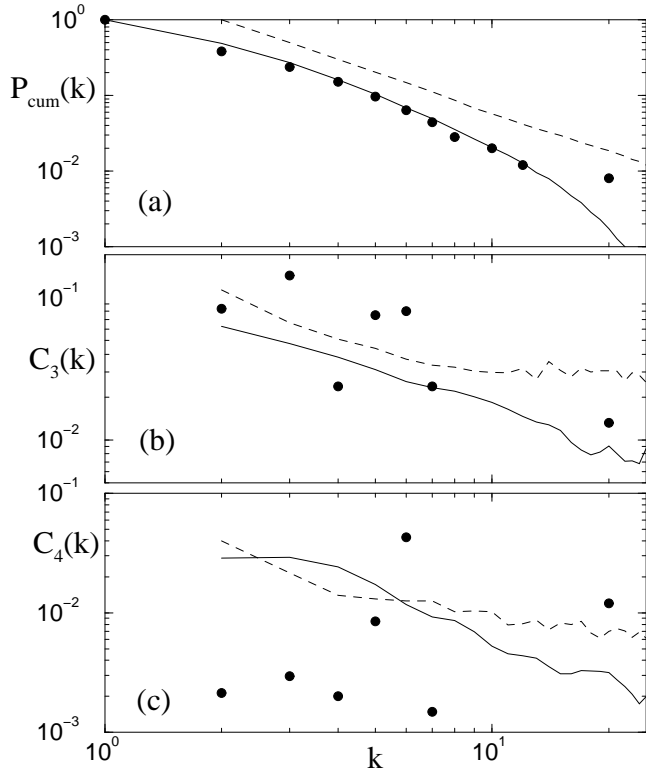}
\end{center}
\caption{\protect 
    Comparing topological features between networks obtained from the
    agent model (solid lines) and random scale-free networks (dashed
    lines), both used to reproduce one real monopartite network of
    sexual contacts (bullets): 
    {\bf (a)} cumulative degree distribution $P_{cum}(k)$,
    {\bf (b)} standard clustering coefficient $C_3(k)$ in
              Eq.~(\ref{c3}) and
    {\bf (c)} clustering coefficient $C_4(k)$ in Eq.~(\ref{c4}).
    Here $N=250$ and samples of $100$ realizations were used.
    For the scale-free network $m=2$ which yields the best
    results for the coefficients (see text).}
\label{fig3}
\end{figure}
\begin{figure}[htb]
\begin{center}
\includegraphics*[width=8.5cm,angle=0]{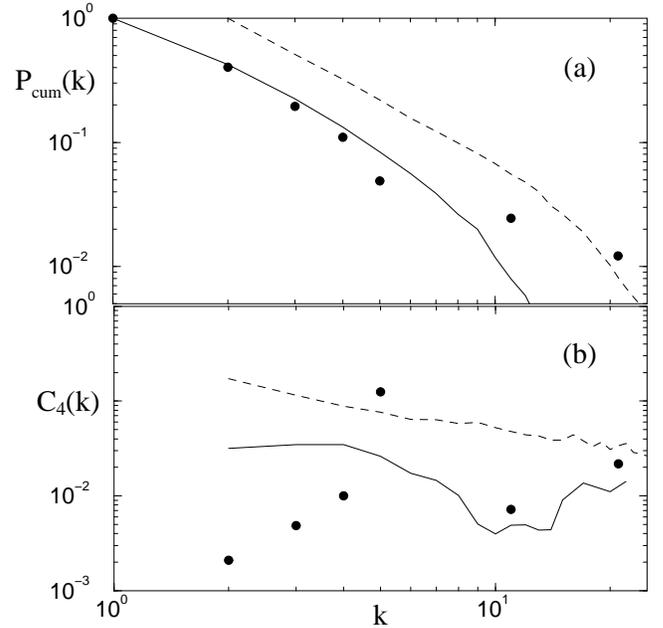}
\end{center}
\caption{\protect
    Distributions for one real bipartite network of sexual contacts
    (bullets) compared with the one of networks obtained from the
    agent model (solid lines) and with random scale-free networks
    (dashed lines): 
    {\bf (a)} cumulative degree distribution $P_{cum}(k)$,
    {\bf (b)} clustering coefficient $C_4(k)$ in Eq.~(\ref{c4}).
    Here $C_3(k)=0$ always, $N=82$, samples of $100$ realizations
    were used for both models and scale-free networks have $m=2$.}
\label{fig4}
\end{figure}

Using the same number of nodes as in the real networks illustrated in
Fig.~\ref{fig2} and considering two types of nodes for the
heterosexual (bipartite) case, we obtain with the agent model similar
results for $L$, $T$, $Q$, $C_3$ and $C_4$, as shown in Table
\ref{tab1} where values represent averages over samples of $100$
realizations. 
Remarkably, for the bipartite case not only the number of
connections and the number of squares are numerically the same, but
also $C_4$ is of the same order of magnitude. 
Similar values of the topological quantities are also obtained for the
monopartite case, with the exception of $C_4$.
Despite this difference, the agent model gives values for the
topological quantities of clustering and cycles much more closely to
the real ones than in random scale-free networks, commonly used to
reproduce such empirical data sets of sexual contacts~\cite{liljeros}.
In Table \ref{tab1} we also show the values obtained for monopartite
and bipartite scale-free networks whose degree distributions are as
close as possible from the distributions of the real networks.

To compare more deeply scale-free networks and networks obtained with
the agent model we study also the distribution of the number $k$ of
sexual partners and the coefficients distributions 
as functions of $k$.
In Fig.~\ref{fig3} we plot these distributions for the monopartite
network of sexual contacts sketched in Fig.~\ref{fig2}b, while in
Fig.~\ref{fig4} we plot the distributions for the bipartite network
(Fig.~\ref{fig2}a). 
In both figures bullets indicate the distributions of the empirical
data, while solid lines indicate the distributions of the networks
obtained with the agent model and dashed lines indicate the
distributions of scale-free networks, with a minimum number of
connections $m=2$, which gives the best fit of a scale-free
distribution to the empirical data with non-zero clustering
coefficient. 
For both models we impose the same size as the real network and take
averages over a sample of $100$ realizations.

As illustrated in Fig.~\ref{fig3}a, the agent model reproduces
accurately the cumulative degree distribution $P_{cum}(k)$ of the real
monopartite network.
For $m=1$ the degree distribution of the scale-free
network yields a better fit to the empirical data than for $m=2$ but
both clustering coefficients are zero for any degree $k$, which is not
realistic as illustrated in Figs.~\ref{fig3}b and \ref{fig3}c.

Figures \ref{fig3}b and \ref{fig3}c show the distribution of
$C_3$ and $C_4$ respectively.
Although the real network (bullets) is very small and therefore
finite size effects appear, one may observe that $C_3$
is larger than $C_4$.
Clearly, both models yield clustering coefficients of the same order of
magnitude as the ones of the real networks, remaining the condition
$C_3(k)>C_4(k)$ observed for the real network.
For scale-free networks the clustering coefficients
are slightly larger than the ones of the agent model. 

Figures \ref{fig4}a and \ref{fig4}b show the cumulative degree
distributions and the distribution of $C_4$ respectively, for the
bipartite network of heterosexual contacts. 
Here, 
$C_3(k)=0$ for all $k$ (not shown).
As one sees, the cumulative distribution for the agent model yields a
better fit to the empirical distribution than the one of scale-free
networks, as seen in Fig.~\ref{fig4}a.
Notice that the cumulative degree distribution for the scale-free
network deviates from a power-law at large values of $k$, since we are
using a bipartite graph which decrease the number of the most connected
nodes. 
Furthermore, comparing Fig.~\ref{fig4}b with Fig.~\ref{fig3}c one
clearly sees that in both cases, $C_4$ has approximately the same shape. 

It is interesting to observe that, while both models reproduce at
least qualitatively the coefficient distributions, in all cases the
agent model fits more accurately the degree distribution of the
empirical data. 
Furthermore, comparing $C_4$ between
homosexual and heterosexual contacts one may rise the hypothesis
that the cliquishness of both types of contacts is similar (see
Sec.~\ref{sec:discuss} below).



In the next Section we will present another application of the
clustering coefficient $C_4$, showing how it can be used to account
for cycles of larger size in any network, and in particular in
bipartite networks. 
\section{Estimating the number of large cycles with squares and triangles}
\label{sec:new}

Recent studies have attracted attention to the cycle structure of complex 
networks, since the presence of cycles has important effects for example
on information propagation through the network~\cite{kim} and on epidemic 
spreading behavior~\cite{petermann}.
In order to avoid numerical algorithms for counting the number of
cycles with arbitrary size which implies long computation times, an
estimate of the fraction of cycles with different sizes was
proposed~\cite{barabasi05}, using the degree distribution $P(k)$ and
the standard cluster coefficient distribution $C_3(k)$.
However, this estimation yields a lower bound for the total number of
cycles and cannot be applied to bipartite networks, as shown below.
In this Section we show that by using both $C_3$ and $C_4$ one is able
to improve that estimation, being suitable at the same time to either 
monopartite and bipartite networks.  
\begin{figure}[b]
\begin{center}
\includegraphics*[width=8.5cm,angle=0]{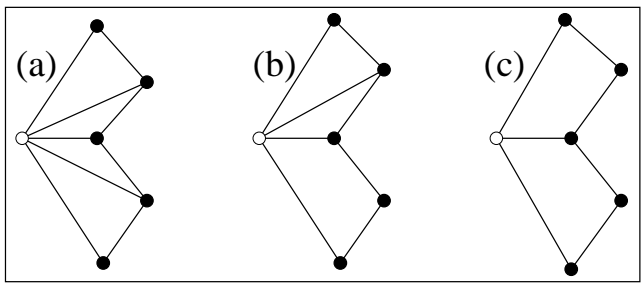}
\end{center}
\caption{\protect
         Illustrative examples of cycles (size $s=6$) where the most
         connected node ($\circ$) is connected to
         {\bf (a)} all the other nodes composing the cycle, forming
         four adjacent triangles.
         In {\bf (b)} the most connected node is connected to all
         other nodes except one, forming two triangles and one
         sub-cycle of size $s=4$, while in
         {\bf (c)} the same cycle $s=6$ encloses two sub-cycles of
         size $s=4$ and no triangles (see text).}
\label{fig6}
\end{figure}

The estimation in Ref.~\cite{barabasi05} considers the set of
cycles with a central node, i.e.~cycles with one node connected to all
other nodes composing the cycle. 
Figure \ref{fig6}a illustrates one of such cycles, where the central
node and each pair of its consecutive neighbors forms a triangle, in a 
total amount of four adjacent triangles.
In such set of cycles, to estimate the number of cycles with size $s$
one looks to the central node of each cycle which has a number, say
$k$, of neighbors.
The number of different possible cycles to occur is 
$n_0(s,k)=\binom{k}{s-1}\tfrac{(s-1)!}{2}$, since
one has $\binom{k}{s-1}$ different groups of $s$ nodes and
in each one of these groups there are $(s-1)!/2$ different ways
in ordering the $s$ nodes into a cycle. 
The fraction of $n_0(s,k)$ of cycles which is expected to occur is
$p_0(s,k)=C_3(k)^{s-2}$, since the probability of having one edge between 
two consecutive neighbors is $C_3(k)$ and one must have $s-2$ edges between
the $s-1$ neighbors.
Therefore, the number of cycles of size $s$ is estimated as
\begin{eqnarray}
N_s &=& Ng_s \sum_{k=s-1}^{k_{max}}
      P(k)n_0(s,k)p_0(s,k) , 
\label{estimate_c3}
\end{eqnarray}
where $P(k)$ is the degree distribution and $g_s$ is a factor which
takes into account the number of repeated cycles.

The estimation in Eq.~(\ref{estimate_c3}) is a lower bound for the
total number of cycles since it considers only cycles with a central
node.
For instance, in Fig.~\ref{fig6}b while cycles of size $s=4$ can be
estimated with Eq.~(\ref{estimate_c3}), the cycle $s=6$ cannot since
it has no central node, and in Fig.~\ref{fig6}c the above equation
cannot estimate any cycle of any size. In fact, Fig.~\ref{fig6}c 
illustrates the type of cycles appearing in bipartite networks, where 
no triangles are observed. For such cycles $C_3(k)=0$ and therefore all 
terms in Eq.~(\ref{estimate_c3}) vanish yielding a wrong estimation of 
the number of cycles.

To take into account cycles without central nodes (Figs.~\ref{fig6}b and 
\ref{fig6}c), one must consider the clustering coefficient $C_4(k)$ 
defined in Eq.~(\ref{c4}). 
One first considers the set of cycles of size $s$ with one node
($\circ$) connected to all the others {\it except} one, as illustrated
in Fig.~\ref{fig6}b. 
In this case, since there are $s-2$ nodes connected to node $\circ$
one has $n_1(s,k)=\binom{k}{s-2}(s-2)!/2$ different possible cycles of
size $s$, with $k$ the number of neighbors of node $\circ$.
The fraction of the $n_1(s,k)$ cycles which is expected to be observed
is given by $p_1(s,k)=C_3(k)^{s-4}C_4(k)(1-C_3(k))$, since
the probability of having $s-4$ connections among
the $s-2$ connected nodes is $C_3(k)^{s-4}$,
the probability that a pair of neighbors of node $\circ$
has to share a common neighbor (different from node
$\circ$) is $C_4(k)$ and the probability that these same pair of 
neighbors have to be not connected is $(1-C_3(k))$.
Writing an equation similar to Eq.~(\ref{estimate_c3}), where instead 
of $n_0(s,k)$ and $p_0(s,k)$ one has $n_1(s,k)$ and $p_1(s,k)$ respectively
and the sum starts at $s-2$ instead of $s-1$, one has an additional number
$N^{\prime}_s$ of estimated cycles which are not considered in estimation
(\ref{estimate_c3}).
Notice that, since for $N^{\prime}_s$ one considers at least one sub-cycle 
of size $s=4$, this additional estimation contributes only for the 
estimation of cycles with size $s\ge 4$. 
We call henceforth sub-cycle, a cycle which is enclosed in a larger
cycle and which do not enclose itself any shorter cycle.

Still, the new estimation $N_s+N_s^{\prime}$ is not suitable for bipartite
networks, since it yields nonzero estimation only for $s=4$. 
To improve the estimation further one must consider not only cycles composed 
by one single sub-cycle of size $s=4$, as done in the previous paragraph, but 
also cycles with any number of sub-cycles of size $s=4$.
Figure \ref{fig6}c illustrates a cycle of size $s=6$ composed by two
sub-cycles of size $4$.
In general, following the same approach as previously, for cycles composed by 
$q$ sub-cycles of size $4$ one finds
$n_q(s,k)=\tfrac{(s-q-1)!}{2}\binom{k}{s-q-1}$ possible cycles of size
$s$ looking from a node with $k$ neighbors and 
a fraction $p_q(s,k)=C_3(k)^{s-2q-2}C_4(k)^{q} (1-C_3(k))^{q}$ of them 
which are expected to be observed.
For $q=0$ one considers cycles as the one illustrated in Fig.~\ref{fig6}a, 
while for $q=1$ and $q=2$ one considers the set of cycles with one and
two sub-cycles with size $4$, as illustrated in Figs.~\ref{fig6}b and
\ref{fig6}c respectively. 
Summing up over $k$ and $q$ yields our final expression
\begin{equation}
N_s = Ng_s \sum_{q=0}^{[s/2]-1}\sum_{k=s-q-1}^{k_{max}}
      P(k)n_q(s,k)p_q(s,k) .
\label{estimate_tot}
\end{equation}
where $[x]$ denotes the integer part of $x$.
In particular, the first term ($q=0$) is the sum in Eq.~(\ref{estimate_c3}).
The upper limit $[s/2]-1$ of the first sum results from the fact that
the exponent of $C_3(k)$ in $p_q(s,k)$ must be non-negative:
$s-2q-2\ge 0$.
The estimation in Eq.~(\ref{estimate_tot}) not only improves the estimated
number computed from Eq.~(\ref{estimate_c3}), but also enables the estimation 
of cycles up to a larger maximal size.
In fact, since in the binomial coefficient $\binom{k}{s-1}$ of 
Eq.~(\ref{estimate_c3}) one must have $s-1\le k\le k_{max}$, 
one only estimates cycles of size up to $k_{max}+1$, while in 
Eq.~(\ref{estimate_tot}) the maximal size is $2k_{max}$, as can be
concluded using both conditions $s-2q-2\ge 0$ and $s-q-1\le k_{max}$.
\begin{figure}[t]
\begin{center}
\includegraphics*[width=8.5cm,angle=0]{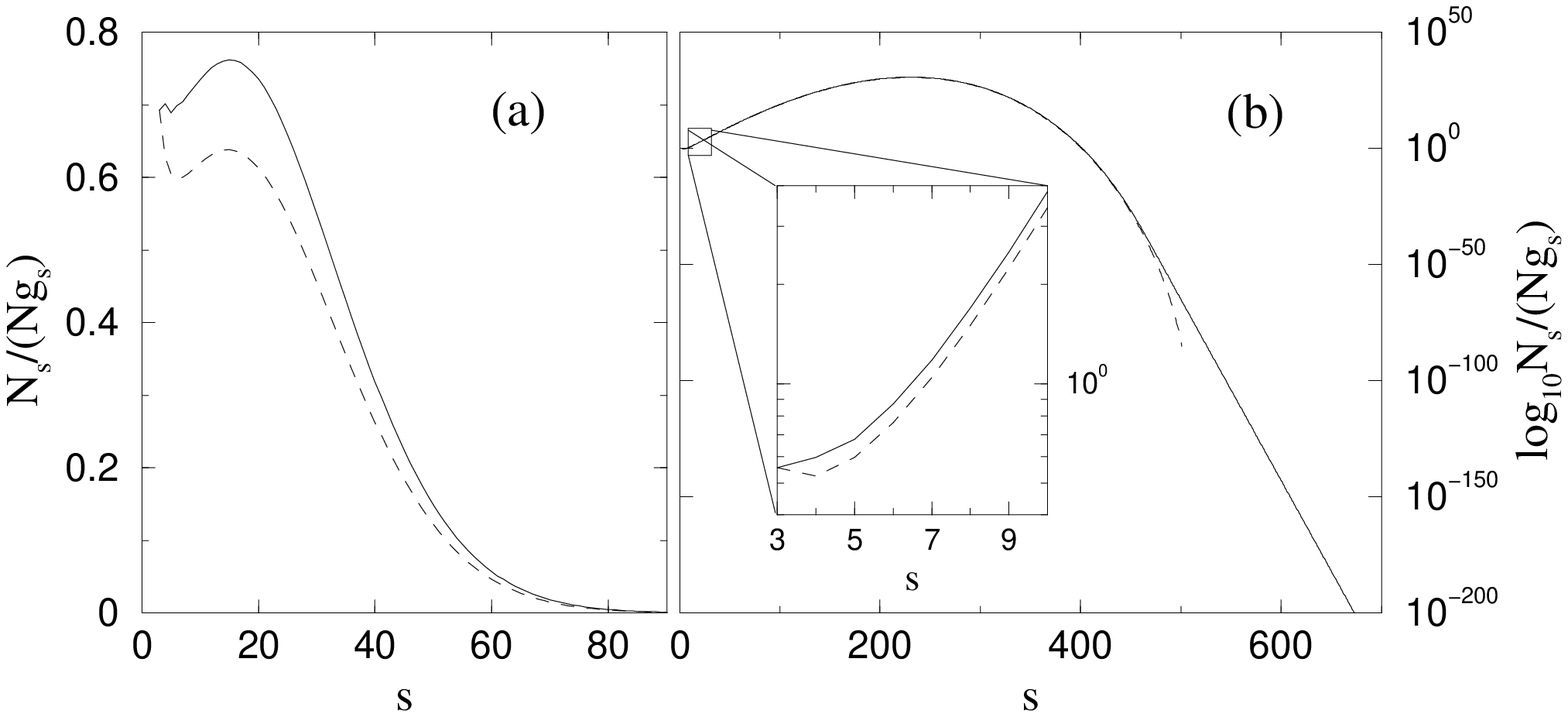}
\end{center}
\caption{\protect
         Estimating the number of cycles using
         Eq.~(\ref{estimate_c3}), dashed lines, and
         Eq.~(\ref{estimate_tot}), solid lines.
         Here we impose a degree distribution $P(k)=P_0 k^{-\gamma}$
         with $P_0=0.737$ and $\gamma=2.5$, and coefficient
         distributions $C_{3,4}(k)=C_{3,4}^{(0)}k^{-\alpha}$ with
         {\bf (a)} $C_3^{(0)}=2$, $C_4^{(0)}=0.33$, $\alpha=0.9$ and
         {\bf (b)} $C_3^{(0)}=1$, $C_4^{(0)}=0.17$, $\alpha=1.1$.
         In all cases $k_{max}=500$.}
\label{fig7}
\end{figure}

Figure \ref{fig7} compares two cases treated in
Ref.~\cite{barabasi05}, both with a degree distribution
$P(k)=P_0k^{-\gamma}$ and coefficient distributions
$C_{3}(k)=C_{3}^{(0)}k^{-\alpha}$, using one value of $\alpha< 1$
(Fig.~\ref{fig7}a) and another one $\alpha> 1$ (Fig.~\ref{fig7}b).
Dashed lines indicate the estimation done with
Eq.~(\ref{estimate_c3}), while solid lines indicate the estimation
done with Eq.~(\ref{estimate_tot}). 
In both cases, the latter estimation is larger.
For $\alpha<1$ the difference between both estimations decreases with
the size $s$ of the cycle.
For $\alpha>1$ the difference between the estimations increases with
$s$ beyond a size $s^{\ast}\lesssim k_{max}$. 
Clearly, from Fig.~\ref{fig7}b one sees that $k_{max}+1$ is the larger
cycle size for which Eq.~(\ref{estimate_c3}) can give an estimation,
while for Eq.~(\ref{estimate_tot}) the estimation proceeds up to
$2k_{max}$ (partially shown).
In both cases, the typical size for which $N_s$ attains a maximum is
numerically the same for both estimations, as expected.
Moreover, for $\alpha>1$ (Fig.~\ref{fig7}b), beyond a size of the
order of $k_{max}$, $N_s/(Ng_s)$ in Eq.~(\ref{estimate_tot}) decreases
exponentially with $s$, and not as a cutoff as observed for
Eq.~(\ref{estimate_c3}). 
In fact, the deviation of Eq.~(\ref{estimate_c3}) from the exponential
tail, is due to the fact that for very large cycle sizes ($s\sim
k_{max}$) Eq.~(\ref{estimate_c3}) can only consider very few terms in
its sum.

Another advantage of the estimation in Eq.~(\ref{estimate_tot}) is that it 
estimates cycles in bipartite networks.
For bipartite network there are no connections between the neighbors,
i.e.~all subgraphs are similar to the one illustrated in
Fig.~\ref{fig6}c.
Therefore all terms in Eq.~(\ref{estimate_tot}) vanish except those
for which the exponent of $C_3(k)$ is zero, i.e.~for $s=2(q+1)$.
Consequently, since $q$ is an integer,
Eq.~(\ref{estimate_tot}) shows clearly that in bipartite networks
there are only cycles of even size, as already known~\cite{holme3}.
Moreover, substituting $q=(s-2)/2$ in Eq.~(\ref{estimate_tot})
yields a simple expression for the number of cycles in bipartite
networks, namely
\begin{equation}
N^{\hbox{\tiny{Bipart}}}_s = Ng_s \sum_{k=s/2}^{k_{max}} P(k) 
           \frac{(s/2)!}{2}\binom{k}{s/2}
           C_4(k)^{s/2-1} .
\label{estimate_bip}
\end{equation}
\begin{figure}[t]
\begin{center}
\includegraphics*[width=8.5cm,angle=0]{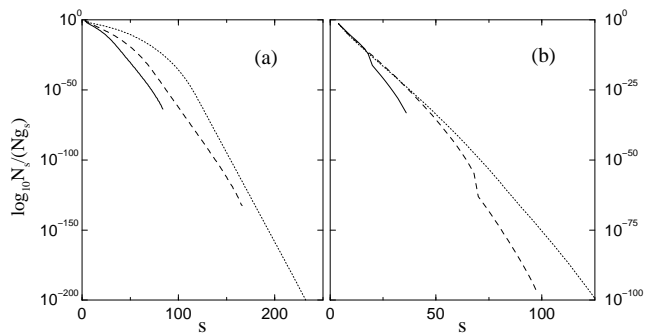}
\end{center}
\caption{\protect
         Estimating the number of cycles for the agent model using
         Eq.~(\ref{estimate_tot}) for
         $N=1000$ (solid lines), $N=5000$ (dashed lines) and
         $N=10000$ (dotted lines) in
         {\bf (a)} a monopartite network and in
         {\bf (b)} a bipartite network, both obtained with the agent
         model.}
\label{fig8}
\end{figure}

Figure \ref{fig8} shows distribution of the fraction $N_s/(Ng_s)$
of cycles as a function of their size $s$, for a monopartite network
(Fig.~\ref{fig8}a) and a bipartite network (Fig.~\ref{fig8}b) composed
by $N=1000,5000$ and $10000$ nodes. This networks were generated from
the agent model described in the previous section.
Here, while monopartite networks show an exponential tail preceded by
a region where the number of cycles is large, bipartite networks are
composed by cycles whose number depend exponentially of their size. 
Furthermore one observes a clear transition for a characteristic size,
which seems to scale with the network size.

It is important to notice that triangles and squares may appear in any
multipartite network (except in bipartite ones, where triangles are
absent).  
Therefore, the estimation described and studied in this Section can be
applied not only to bipartite networks but to any multipartite network
of any order.

\section{Discussion and conclusions}
\label{sec:discuss}

We introduced a clustering coefficient similar to the standard one,
which instead of measuring the fraction of triangles in a network
measures the fraction of squares, and showed that with this clustering
coefficient it is also possible to characterize topological features
in complex networks, usually done with the standard coefficient.
We showed explicitly that the range of values of the probability to
acquire long-range connections in small-world networks and the typical
clustering coefficient distributions of either random scale-free and
hierarchical networks are approximately the same.
In addition, we showed that this second clustering coefficient enables
one to quantify the cliquishness in bipartite networks where triangles
are absent. 
Thus, one should take triangles and squares simultaneously
as the two basic cycle units in any network.

An application of both clustering coefficients was
proposed, namely to estimate the number of cycles in any network,
either monopartite or multipartite.
Using a recent estimation which yields a lower bound of the number of
cycles in monopartite network up to a size $s<k_{max}+1$ where $k_{max}$
is the maximum number of neighbors in the network, we deduce a more
general expression which not only improves the previous estimation but
is also suitable for bipartite networks and enables one to estimate
cycles of size up to $2k_{max}$.
Furthermore, in the particular case of bipartite networks our
estimation yields as a natural consequence that only cycles of even
size may appear. 

We also studied a concrete example of two sexual networks, one
where only heterosexual contacts occur (bipartite network) and another
with homosexual contacts (monopartite).
The results obtained with the two real networks were compared with the
ones obtained with a scale-free network and with an agent model
recently introduced.
Our results emphasize that, in general, the agent model seems to be
more suitable to reproduce networks of sexual contacts than the
standard approach with random scale-free networks.
Furthermore, our results for the clustering distribution of both real
sexual networks gave some evidence that the clustering
ability in sexual networks probably does not depend on the type of sexual
contact (homosexual or heterosexual).
To strengthen this hypothesis it is necessary to use larger networks of
sexual contacts and apply the topological quantities here described.
These and other questions are being analyzed in more detail and will
be presented elsewhere.

\section*{Acknowledgments}

MCG thanks Deutscher Akademischer Austausch Dienst (DAAD), Germany,
and PGL thanks Funda\c{c}\~ao para a Ci\^encia e a Tecnologia (FCT),
Portugal, for financial support. 



\end{document}